\documentclass[conference]{IEEEtran}
\IEEEoverridecommandlockouts
\usepackage{cite}
\usepackage{amsmath,amssymb,amsfonts}
\usepackage{algorithmic}
\usepackage{graphicx}
\usepackage{textcomp}
\usepackage{xcolor}
\usepackage{multirow}
\usepackage{array} 
\usepackage[ruled,vlined]{algorithm2e}
\def\BibTeX{{\rm B\kern-.05em{\sc i\kern-.025em b}\kern-.08em
    T\kern-.1667em\lower.7ex\hbox{E}\kern-.125emX}}
\begin{document}

\title{Ultrasound Image Segmentation of Thyroid Nodule via Latent Semantic Feature Co-Registration\\
\thanks{$\ast$Corresponding author. $\dagger$These authors contributed equally to this work.}
}

\author{\IEEEauthorblockN{Xuewei Li$\rm ^{a,b,c,d\dagger}$, Yaqiao Zhu$\rm ^{b,c,d\dagger}$, Jie Gao$\rm ^{a,b,c}$, Xi Wei$\rm ^{e}$, Ruixuan Zhang$\rm ^{a,b,c}$, Yuan Tian$\rm ^{a,b,c}$, and Zhiqiang Liu$\rm ^{a,b,c\ast}$}
\IEEEauthorblockA{
\textit{$^a$College of Intelligence and Computing, Tianjin University}\\
\textit{$^b$Tianjin Key Laboratory of Cognitive Computing and Application}\\
\textit{$^c$Tianjin Key Laboratory of Advanced Networking}\\
\textit{$^d$School of Future Technology, Tianjin University}\\
\textit{$^e$Department of Diagnostic and Therapeutic Ultrasonography, Tianjin Medical University Cancer Institute and Hospital}\\
Tianjin, China}
Email: \{lixuewei, zhuyaqiao, gaojie\}@tju.edu.cn, weixi@tmu.edu.cn, \{zrx\_6566, tiany, tjubeisong\}@tju.edu.cn
}

\maketitle

\begin{abstract}
Segmentation of nodules in thyroid ultrasound imaging plays a crucial role in the detection and treatment of thyroid cancer. However, owing to the diversity of scanner vendors and imaging protocols in different hospitals, the automatic segmentation model, which has already demonstrated expert-level accuracy in the field of ultrasound imaging segmentation, finds its accuracy reduced as the result of its weak generalization performance when being applied in clinically realistic environments. To address this issue, the present paper proposes ASTN, a novel framework for accurate and generalizable segmentation of thyroid nodules. ASTN incorporates a unique co-registration module that concentrates on the lesion area. This module enhances segmentation accuracy by using anatomical structural information from various datasets. In particular, prior to co-registration, an encoder is used to extract the latent semantic features of the lesion area in both the target image and atlas (a template set composed of multiple real anatomical structures). The features are thereafter registered subsequent to their combination with one another. For the label fusion, we estimate the similarity of results at different stages, adaptively allocate weights to each. This strategy aims to improve the model's fault tolerance and segmentation accuracy in different data domains. Additionally, this paper also provides an atlas selection algorithm to retain more priori information and alleviate the challenges associated with co-registration. Thanks to the method we proposed, as shown by the evaluation results collected from the datasets of different devices, the model generalization has been greatly improved while maintaining a high level of segmentation accuracy.
\end{abstract}


\section{Introduction}
Over the past decade, the incidence rate of malignant thyroid nodules has escalated by 65\%, ultrasound has become the preferred method for evaluating thyroid nodules in diagnosis \cite{li2023fusing}, and the precise segmentation of nodules in thyroid ultrasound imaging has turned into a vital step in the detection and treatment of thyroid cancer. Considering the heavy reliance on the experiences and expertise of clinical practitioners in diagnosing thyroid nodules through ultrasound imaging \cite{yuan2022csm}, a universal and accurate computer-aided diagnostic system is essential to prevent misdiagnosis. One of the key challenges faced in the development of such systems is the variability of training datasets, influenced by differences in scanner vendors and imaging protocols. This variability often hampers the performance of segmentation models, particularly when they are applied to data from different domains than those they were trained on. As evidenced in studies led by Z. Su \cite{su2023rethinking} and J. De Fauw \cite{de2018clinically}, models achieving high accuracy in segmenting images from a single device often experience more than a twofold increase in error rate when applied to images from multiple devices. This poor generalization capacity has emerged as a fundamental hurdle in applying deep-learning models in clinical settings \cite{zhang2020generalizing}.

\begin{figure}[t]
\centerline{\includegraphics[scale=0.16]{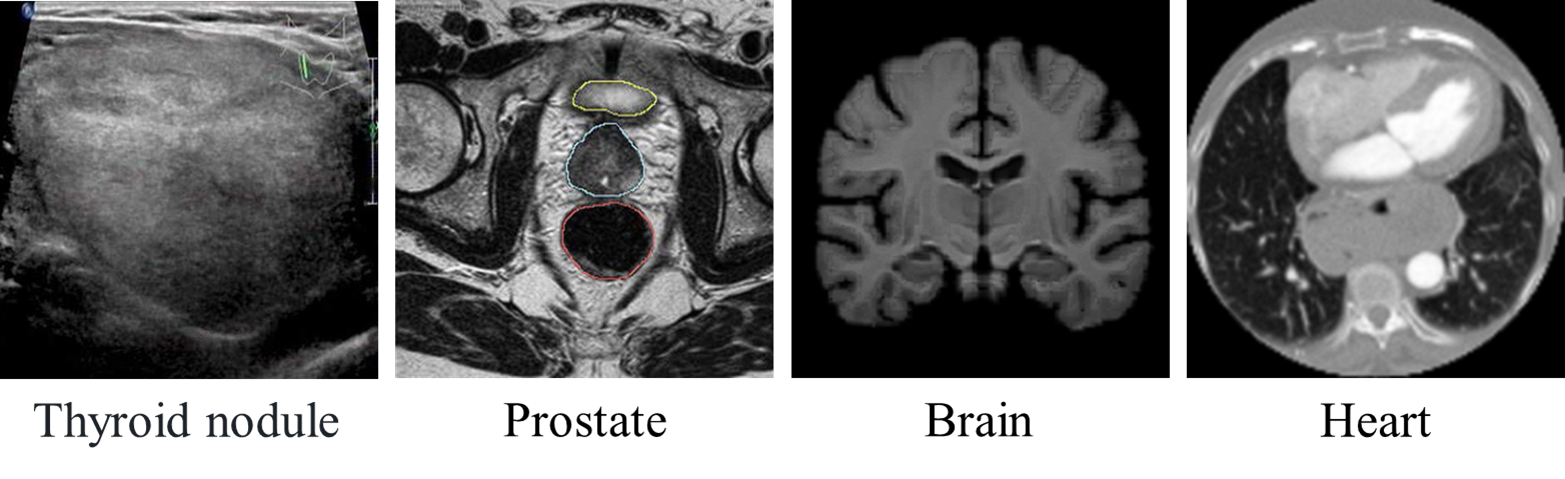}}
\caption{\textbf{Comparison of Thyroid Nodule with Other Body Parts}}
\label{first}
\end{figure}

In response to the challenge of improving generalization capacity, several research teams have explored domain adaptation or domain generalization strategies, utilizing data manipulation and adversarial training to address domain shifts\cite{wang2022generalizing, chen2020unsupervised, chen2019joint}.
However, these methods often come with a significant investment in terms of human resources and time, both for data collection and model training. Prior to the emergence of powerful convolutional networks used for pixel-level segmentation, the primary approach in segmenting the biomedical images with known structure is the conventional co-registration-based method\cite{sinclair2022atlas,isgum2009multi}, which aligns known anatomical topological structures with target images by using co-registration in order to achieve segmentation, thus requiring less computational resources compared to domain generalization approaches. Along with the proposal of Spatial Transformer Networks (STN)\cite{jaderberg2015spatial}, the traditional co-registration process can be realized by using deep networks. Compared with the networks that perform pixel-level segmentation\cite{chen2020banet}, networks that employ topological structure co-registration for segmentation are believed to exhibit superior generalization, particularly when handling data from various scanner vendors and imaging protocols \cite{xie2023deep}. This assertion forms the basis of our study, which aims to leverage the strengths of topological co-registration within a deep learning context to address the challenges of domain variability in ultrasound image segmentation.

In recent years, there have been constant endeavors to improve the co-registration-based medical imaging segmentation. For instance, Bin Huang\cite{huang20213d} developed a lightweight neural network that improved the accuracy of segmenting head and neck cancer lesions in organ-at-risk delineation through co-registration techniques. Similarly, Jiazhen Zhang's work \cite{zhang2022atlas}, which utilized an anatomical atlas for post-processing segmentation results, demonstrated the potential for enhancing image segmentation accuracy by dynamically combining atlas images with semantic segmentation outcomes. In order to mitigate the false pixel predictions caused by noise during cardiac segmentation, Atlas-ISTN\cite{sinclair2022atlas} directly co-registered the constructed atlas labels with the segmented results after obtaining an initial cardiac segmentation mask. However, these advances encounter unique challenges when applied to thyroid nodule segmentation in ultrasound images. Unlike the body parts mentioned above, thyroid nodules exhibit minimal textural contrast with surrounding tissues, as illustrated in Fig.~\ref{first}. This similarity in texture has led to lower accuracy in previous co-registration attempts for thyroid nodules. This specific challenge underscores the need for a more refined approach in ultrasound image segmentation for thyroid nodules, which is the focus of our study.

\begin{table}[htbp]
\renewcommand\arraystretch {1.25}
\setlength{\tabcolsep}{5mm}{
\centering
\caption{Comparison of the DSC results of the co-registration of different data including Thyroid Nodules(TN), Left Ventricular Myocardium (LVM), Left Atrial Cavity (LAC), Ascending Aorta (AA), Liver, Optic Cross (OC), Lumbar Vertebrae (LV), and Left Ventricular Cavity (LVC) using the SyN method}
\label{syn}
\begin{tabular}{c|cccc}
\hline
\textbf{Data} & \textbf{TN} & \textbf{LVM} & \textbf{LAC} & \textbf{AA}             \\ 
DSC↑              & \textbf{27.3}            & 52.6                   & 64.49                                 & 45.55                                    \\ \hline
\textbf{Data} & \textbf{Liver}           & \textbf{OC}  & \textbf{LV}             & \textbf{LVC} \\ 
DSC↑              & 75.13                    & 55.7                   & 74                                    & 70.79                                    \\ \hline
\end{tabular} }
\end{table}

In this section, we conducted tests on the thyroid nodules co-registration using symmetric image normalization (SyN) \cite{avants2008symmetric}, which is regarded as one of the top-performing co-registration methods. Surprisingly, the co-registration Dice Similarity Coefficient (DSC) for thyroid nodules was found to be only 27.3\%, significantly lower than for other body parts shown in TABLE~\ref{syn} \cite{ding2022cross, huang20213d, ding2020cross, xie2023deep}. This stark contrast underscores the unique challenges posed by thyroid nodules, primarily due to their subtle textural differences from surrounding tissues and their unpredictable growth positions within ultrasound images. Current methods \cite{wu2023latent, qiu2021rsegnet}, while effective in aligning the overall image shape, often overlook the critical semantic details of the nodules, leading to a loss of original anatomical integrity. Furthermore, existing atlas selection techniques \cite{van2022deep,zhang2021deep} fall short in ensuring spatial proximity between the lesion regions in the atlas and the target images, thus affecting the accuracy of co-registration and subsequent segmentation \cite{ding2022aladdin}.

To address these challenges, we propose the co-registration and segmentation framework ASTN, specifically tailored for thyroid nodules in ultrasound images. Our approach introduces two innovative components: a novel atlas selection algorithm and a dictionary system. The atlas selection algorithm, guided by the Regional Correlation Score (RCS), focuses on constructing dictionary elements by selecting regionally representative images and labels. This approach aims to bridge the spatial and textural gaps between the atlas and the target lesion areas, thereby simplifying the co-registration process. The dictionary system, comprising a newly developed co-registration network termed 'half-STN' and an advanced label fusion method, leverages the semantic information of the nodules. It determines the spatial relationship between the dictionary elements and the nodules in the target images. This system marks a significant shift in co-registration strategy, moving from traditional alignment of superficial image features to a deeper alignment of latent semantic features. By doing so, it generates a Displacement Field (DF) that maintains the integrity of the nodule areas, thus reducing the segmentation discrepancies caused by variations in imaging devices, with the aim of enhancing the generalization of the model. In the present study, the experiments were conducted using datasets from Thyroid Ultrasound Image (TUI) collected by two different devices. The key contributions of this research are summarized as follows: 

\begin{itemize}
    \item We introduce a pioneering co-registration-based segmentation method tailored for thyroid nodules in ultrasound images. This method uniquely extracts and utilizes semantic information during the co-registration process, ensuring the preservation of critical anatomical structures for more accurate segmentation masks. Additionally, we refine a method for calculating weights during the label fusion of the multi-stage warped atlas.
    \item Our work proposes an atlas selection algorithm. This algorithm is designed to ensure that the atlas's data distribution comprehensively encompasses the variability observed in the target images, thereby significantly boosting the co-registration's effectiveness.
    \item We evaluate the performance of our methodology on the TUI dataset, achieving remarkable results. The DSC for co-registration has been elevated to 88.59\%, and the IoU of the segmentation results increased by 1.34\% and 6.524\% in the known and unseen domains compared to existing methods.
\end{itemize}

\section{Methodology}
\begin{figure*}[htbp]
\centering 
\includegraphics[width=1\textwidth]{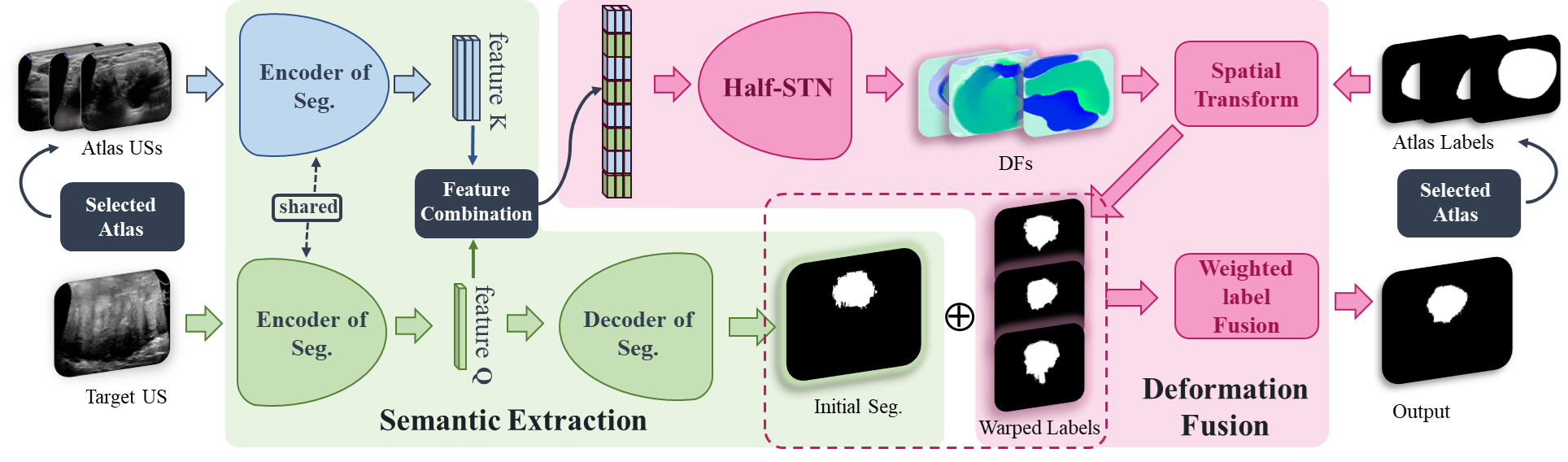} 
\caption{\textbf{Overview of our ASTN.} ASTN encompasses the components of Atlas Selection, Semantic Extraction, and Deformation Fusion. During the training, a fixed selected atlas is inputted into the network along with the target ultrasound image(target US). The Deformation Fusion module leverages the semantic features and the initial segmentation result provided by Semantic Extraction to generate the final segmentation. The Feature Combination and Weighted label Fusion are depicted in Fig.~\ref{feature} and Fig.~\ref{fuse}, respectively.}
\label{ASTN}
\end{figure*}

In this section, we detail the implementation methodology of ASTN, illustrated in Fig. \ref{ASTN}. The framework comprises two primary components: a novel atlas selection algorithm and a dictionary system. The atlas selection algorithm is integral for constructing the dictionary atlas, which forms the foundation for our targeted segmentation approach. Meanwhile, the dictionary system plays a pivotal role in the segmentation process, consisting of two specialized components. The first, the Semantic Extraction part (highlighted in green in the figure), is dedicated to accurately localizing nodule features within the ultrasound images. The second component, the Deformation Fusion part (highlighted in red), is responsible for warping and integrating the atlas labels based on the features. 

\begin{figure}[t]
\centerline{\includegraphics[scale=0.24]{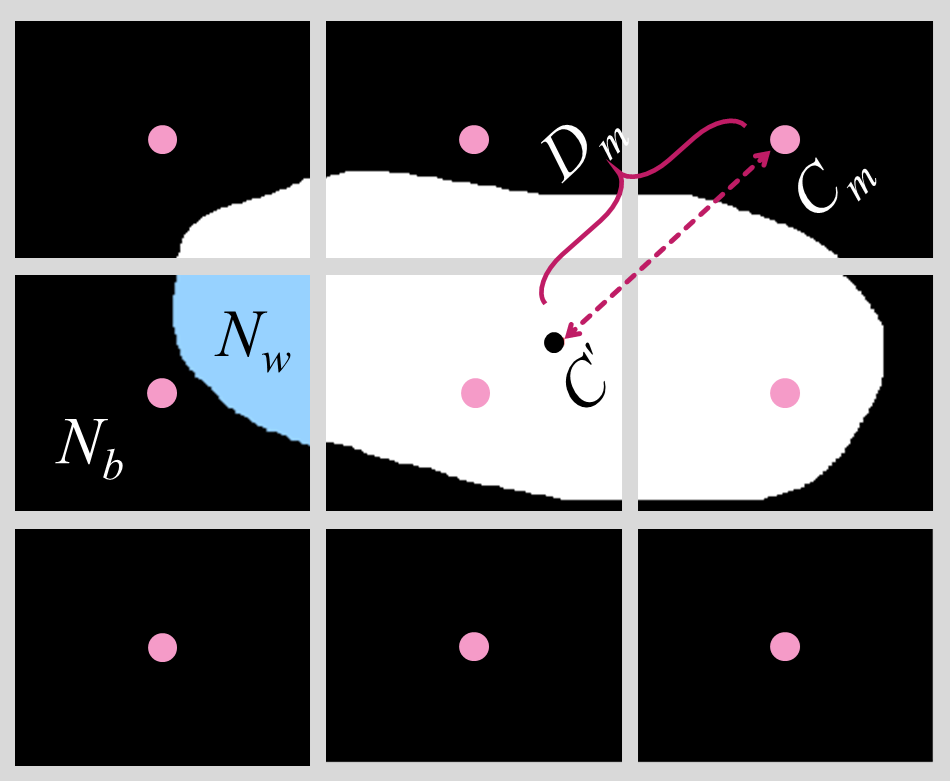}}
\caption{\textbf{Regional Correlation Score.} The picture above illustrates the RCS computation process when $M$ is 9. The red dots represent the centroids $C_m$ of the $m$ region, and the black dot represents the centroid $C^{\prime}$ of the entire nodule. The number of nodule pixels in the current region is $N_w$, and the number of background pixels is $N_b$}
\label{RCS}
\end{figure}

\subsection{Regional Correlation Score}

Addressing the complexities inherent in co-registering thyroid nodules in ultrasound imaging, our methodology confronts the challenges posed by the variability of imaging positions and the sparse distribution of nodules. To mitigate the impact of these challenges, our atlas selection algorithm strategically utilizes the RCS to assemble an atlas that effectively represents thyroid ultrasound images with a uniform distribution of nodules.

The novel atlas selection algorithm calculates the RCS for each candidate image. To do this, we first dissect each candidate image into a grid of regions, denoted as $u$ rows by $v$ columns, resulting in $u \times v = M$ total regions, as illustrated in Fig.~\ref{RCS}. For each region in a given candidate (labelled as $a$), the RCS computation is carried out as follows:

\textbf{Standard Deviation of Proportion} ($Std(P_m)$): This is calculated as 
\begin{align}
    P_m = \frac{N_w}{N_w + N_b}
\end{align}    
where $N_w$ represents the number of pixels in the nodule within region $m$, and $N_b$ denotes the background pixels in the same region. $P_m$ thus signifies the proportion of the nodule area within that specific region.

\textbf{Standard Deviation of Centroid Distance} ($Std(D_m)$): Here, 
\begin{align}
    D_m = \sqrt{(C_x^{\prime}  - C_{mx})^2 + (C_y^{\prime} - C_{my})^2}
\end{align}
which calculates the Euclidean distance between the centroid of the nodule area and the centroid of region $m$. The coordinates of these centroids are denoted as $(C_{mx}, C_{my})$ for the region and $(C_x^{\prime}, C_y^{\prime})$ for the nodule area. This distance metric helps in understanding how centrally located the nodule is within each region.

The RCS for each region ($s_m$) is then defined by the formula: 
\begin{align}
s_m = Std(P_m) - Std(D_m)
\end{align}
which captures both the spatial distribution and the centrality of the nodules within each region, providing a comprehensive score. Higher scores indicate a stronger correlation between the candidate image and the target region, suggesting a more accurate alignment potential with nodules in that region.

Once the RCS for all candidates and regions has been computed, we aggregate these scores to form the atlas dictionary. The atlas is then assembled by selecting the candidate with the highest RCS for each region, yielding the final atlas as: 
\begin{align}
Atlas = \{argmax(S_{m}), m = 1, 2, \cdots, M\}
\end{align}

\begin{figure}[t]
\centerline{\includegraphics[scale=0.28]{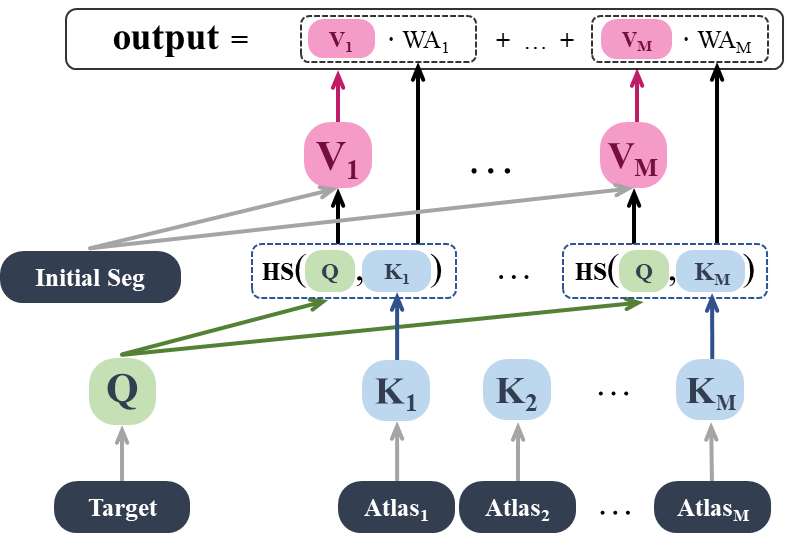}}
\caption{\textbf{Dictionary System}. Half-STN denoted as HS, WA stands for warped atlas}
\label{dsystem}
\end{figure}

\subsection{Dictionary System}

\begin{figure}[t]
    \centering
    \includegraphics[scale=0.26]{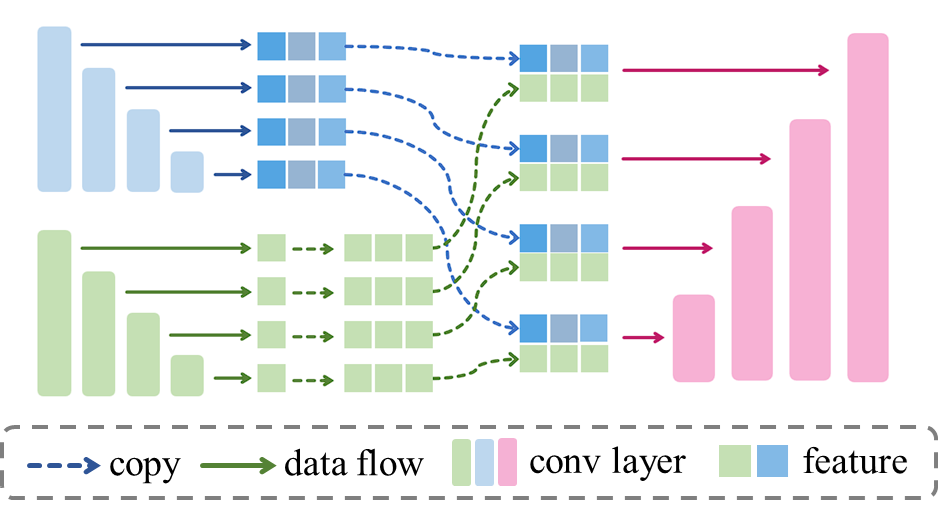}
    \caption{\textbf{Feature Combination.} The upper blue encoder receives $I_A$, generating features of dimensions $M\times N$. The lower green encoder receives the $I_T$ and obtains features of dimensions $1\times N$. After combining the two into a $M\times N$ dimensional feature, it is used as an input to the red half-STN}
    \label{feature}
\end{figure}

\begin{algorithm}[th]
	\caption{The Algorithm of Dictionary System}
    \renewcommand{\algorithmicrequire}{\textbf{Input:}}
    \renewcommand{\algorithmicensure}{\textbf{Output:}}
	\begin{algorithmic}[1]
		\REQUIRE Training dataset $\mathcal{D_T}$; Atlas $\mathcal{A}$; Batch Size $\mathcal{B}$ 
		\ENSURE Optimal \textit{Enc, Dec, HS}   
		\STATE $\Theta_{Enc},\Theta_{Dec},\Theta_{HS},  \leftarrow $initialize network randomly \tcp{Initialize the encoder and decoder of seg network parameters and Half-STN parameters} 

	\FOR {$iter\ in\ range(SegEpoch)$}
            
		  \STATE Sample a mini-batch $ (X, Y) = \{(x_i, y_i)\}^{|\mathcal{B}|}_{i=1}$ from  $\mathcal{D_T}$
            \STATE $ F \leftarrow Enc(X) $ \tcp{Feature of \textit{X}}
            \STATE $ \tilde{y}_0 \leftarrow Dec(F) $ \tcp{Initial seg result}
            \STATE $ \mathcal{L}_{seg} \leftarrow (\Vert \tilde{y} - Y \Vert_2 )^2 $ \tcp{Calculate seg loss}
            
            \tcc{Update seg network parameters}
            \STATE $\Theta_{Enc} \leftarrow \Theta_{Enc} - \nabla_{\Theta_{Enc}}\mathcal{L}_{seg}$ 
            \STATE $\Theta_{Dec} \leftarrow \Theta_{Dec} - \nabla_{\Theta_{Dec}}\mathcal{L}_{seg}$ 
        \ENDFOR
        \FOR{$iter\ in\ range(RegEpoch) $}
            \STATE Sample a mini-batch $ (X, Y) = \{(x_i, y_i)\}^{|\mathcal{B}|}_{i=1}$ from  $\mathcal{D_T}$
            \STATE Get atlas images and labels $(X_A,Y_A) $ from $\mathcal{A}$
            \STATE $F, F_A \leftarrow Enc(X, X_A)$
            \STATE $\tilde{y}_0 \leftarrow Dec(F)$
            \STATE $ \mathcal{L}_{seg} \leftarrow (\Vert \tilde{y} - Y \Vert_2 )^2 $ \tcp{Calculate seg loss}
            \STATE $ W \leftarrow HS(F,F_A)$ \tcp{Get warped labels}
            \STATE Compute $V$ using Eq. \eqref{eq1}
            \STATE $\tilde{y} \leftarrow \sum V \cdot W$ final prediction
            \STATE Compute $\mathcal{L}_{reg}$ using Eq. \eqref{eq3}\\
            \tcc{Update all network parameters}
            \STATE $\Theta_{Enc} \leftarrow \Theta_{Enc} - \nabla_{\Theta_{Enc}}(\mathcal{L}_{seg} + \lambda\mathcal{L}_{reg})$
            \STATE $\Theta_{Dec} \leftarrow \Theta_{Dec} - \nabla_{\Theta_{Dec}}\mathcal{L}_{seg}$ 
            \STATE $\Theta_{HS} \leftarrow \Theta_{HS} - \nabla_{\Theta_{Enc}}(\lambda\mathcal{L}_{seg} + \mathcal{L}_{reg})$
        \ENDFOR
         
	\end{algorithmic}
	\label{alg1}
\end{algorithm}

\paragraph{Half-STN Model} To maintain anatomical integrity during co-registration, we propose the Half-STN (HS) model, which utilizes semantic information of ultrasound images derived from a segmentation network to generate DF. Compared to the DF obtained from the entire image using STN, in-depth semantic information avoids blind pixel-level alignment that can compromise topology structure, resulting in more faithful deformation outcomes. A representative segmentation network Unet is used as the backbone in this section to facilitate description. 

Since an initial segmentation result is needed in the subsequent label fusion stage to assign an appropriate weight, denoted as $V$ in Fig.~\ref{dsystem}, to each warped atlas, the Unet is first trained for segmentation independently before building the whole framework. This process enables the encoder of Unet to extract semantic features of nodules from ultrasound images, which serve as the information source for generating the DF in the HS. During performing co-registration, it is typically necessary to input both the fixed and moving images, which corresponds to the target feature $Q$ and the atlas feature $K$ in our HS method. The manner in which semantic features are obtained is as follows:
\begin{align}
    Q = U_{enc}(I_T),\quad K_A = U_{enc}(I_A)
\end{align}
where $U_{enc}$ denotes the encoder of Unet, $I_T$ denotes the target image, there are $M$ images in the atlas $I_A$, whose features are defined as $K_A=\{f_a,a=1,2,\cdots,M\}$. In the subsequent training, it is essential to maintain the feature extraction capacity of $U_{P_1}$ and assure the rationality of the initial segmentation result, so the following loss function will serve as an item of the overall loss of ASTN, enabling Unet to actively engage in the holistic optimization of the model:
\begin{equation}
\begin{gathered}
\mathcal{L}_{sim}(Seg_{inital},L_T)=(\Vert Seg_{inital}-L_T\Vert_2 )^2 \\
Seg_{inital}=U_{dec}(Q)
\end{gathered}
\end{equation}
where $U_{dec}$ denotes the decoder of Unet, $L_T$ denotes the target label, and mean square error (MSE) is used to compute the segmentation loss of Unet $\mathcal{L}_{sim}$.

In order for the half-STN to capture the correspondence between the fixed image and moving image, the combination of $Q$ and $K$ is performed by broadcasting the dimensions and utilizing the concatenation method as depicted in Fig.~\ref{feature}. Additionally, to preserve the high-dimensional information, the information propagated through each skip connection requires individual concatenation.

To discern the spatial positional relationship between the implicit features of nodules extracted from $Q$ and $K_A$, we have devised a Half-STN network that is capable of capturing the DF through the features. When constructing the Half-STN, we employe a network architecture similar to the decoder structure in Semantic Extraction, with the aim of enhancing the universality of the ASTN framework while enabling comparison of co-registration performance across different structural decoders. The co-registration process is outlined as follows:
\begin{equation}
\begin{gathered}
\widetilde{L}_A=L_A\circ (Id+DF) \\
DF_A=HS(Q,K_A)
\end{gathered}
\end{equation}
where $DF_A$ is the DF of $I_A$ towards $L_T$, wherein the size matches that of the original image and encompasses the coordinates of each pixel in the original image after the variations. $\widetilde{L}_A$ denotes the warped atlas labels, while $Id$ represents the identity transform \cite{bajcsy1989multiresolution}. At this stage, the framework incorporates the following loss function to optimize the Half-STN:
\begin{align} \label{eq3}
\mathcal{L}_{reg}(\widetilde{L}_A,L_T,DF)=\mathcal{L}_{sim}(\widetilde{L}_A,L_T)+\lambda_1\mathcal{L}_{sot}(DF)
\end{align}
where $\mathcal{L}_{sim}$ is equivalent to the loss function used in segmentation. When $\widetilde{L}_A$ and $L_T$ are excessively similar, the $DF$ employed for co-registration loses its smoothness, which is impractical in biological anatomy. Hence, we restrict $DF$ to maintain its smoothness by utilizing the diffusion regularizer $\mathcal{L}_{sot}$ on the displacement field's planar gradient. The coefficient $\lambda_1$ signifies the smoothness loss factor. Controlling for the effect of segmentation loss with $\lambda_2$, the final loss $\mathcal{L}$ for the whole network is:
\begin{align} \label{eq2}
\mathcal{L} = \mathcal{L}_{reg}(\widetilde{L}_A,L_T,DF)+\lambda_2\mathcal{L}_{sim}(Seg_{inital},L_T)
\end{align}

\begin{figure}[t]
    \centering
    \includegraphics[scale=0.26]{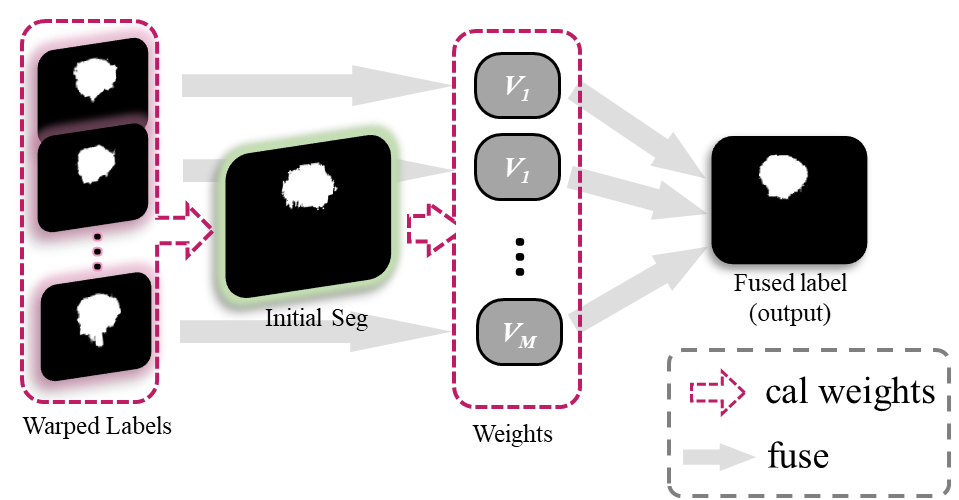}
    \caption{\textbf{Warped label fusion.} The initial segmentation is evaluated against each warped label using DSC, yielding respective weights upon normalization as indicated by the red arrow. The fusion process, depicted by the grey arrow, involves the weighted summation of warped labels to obtain the final segmentation outcome.}
    \label{fuse}
\end{figure}

\begin{figure*}[t]
\centerline{\includegraphics[scale=0.44]{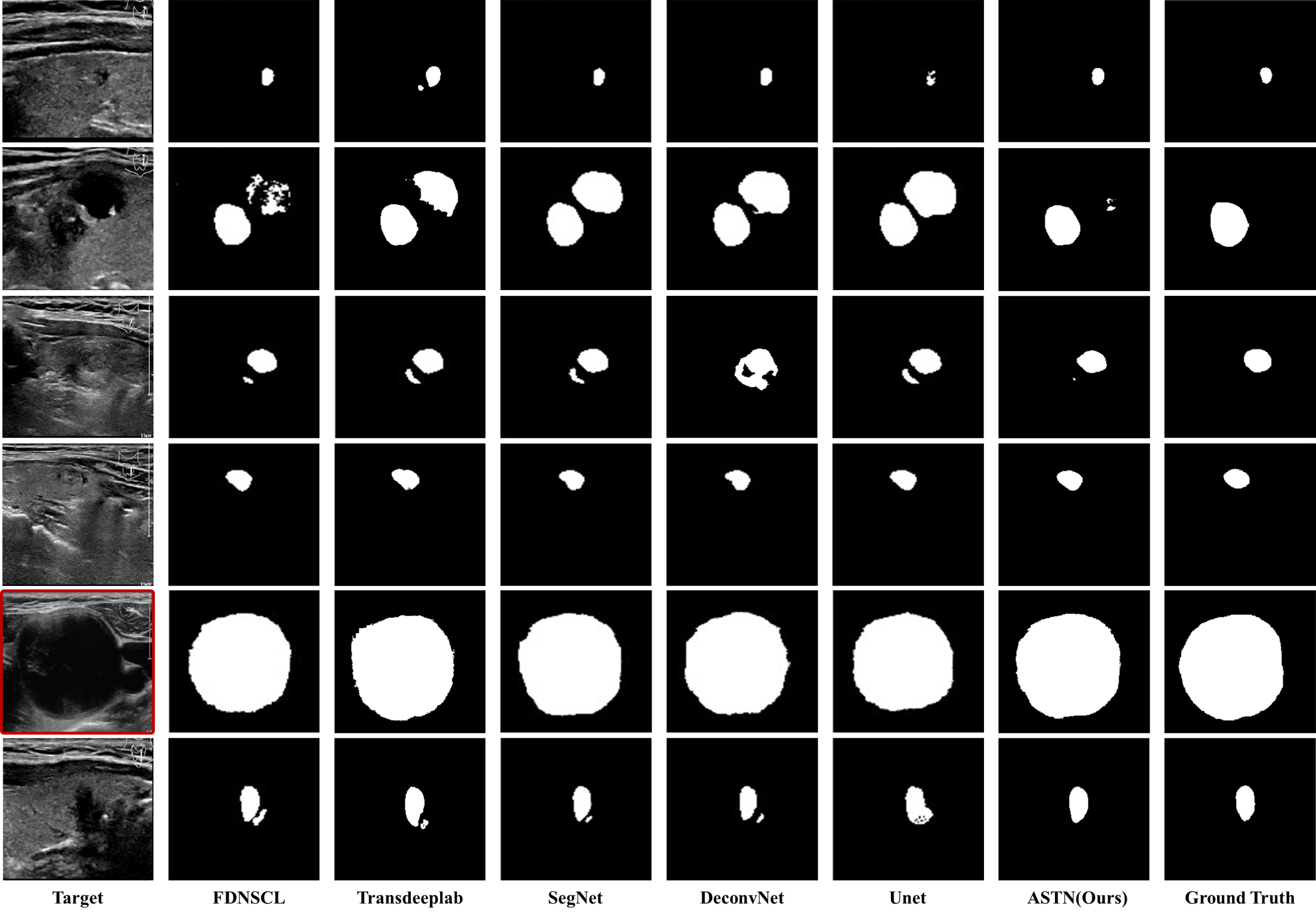}}
\caption{\textbf{Comparison Experiment.} The first column shows the target ultrasound images, while the last column shows the annotated labels made by experts. The remaining columns exhibit the forecasting outcomes of different models. Nodule areas are depicted in white. The co-registration process of the red target is illustrated in Fig.~\ref{register}.}
\label{bigfig}
\end{figure*}


\paragraph{Warped label fusion} To effectively address the challenges posed by the dispersed distribution of nodules in the atlas and the resultant inconsistencies in the precision of warped labels obtained through co-registration, we have developed an innovative label fusion strategy. This approach is designed to enhance the fault tolerance of the co-registration process. It ensures that the final segmentation mask remains accurate, even in the presence of co-registration errors in individual atlas elements. The strategy, including its application to the initial segmentation as a reference, is detailed in Fig.~\ref{fuse}.


In our novel approach to assigning weights to each warped label, we diverge from traditional methods that rely heavily on the information itself, as done by S. K. Warfield \cite{warfield2004simultaneous}, or the creation of a subnetwork as suggested by Long Xie \cite{xie2023deep}. Instead, we adopt a more streamlined method that maintains accuracy while minimizing the number of required parameters. This is achieved by leveraging the initial segmentation ($Seg_{initial}$) obtained from the Semantic Extraction process. The DSC is calculated between the warped label $\widetilde{L}_A=\{\tilde{l}_a,a=1,2,\cdots,M\}$ and $Seg_{initial}$, and after normalization, obtain the weights $v_a$ for the current warped label. In practice, in order to fully exploit the target image information, the fusion approach incorporates $Seg_{initial}$ in the final weighted summation. Additionally, a weight of $v_0=\frac{1}{M+1}$ is assigned to $Seg_{initial}$. The expression for $v_a$ is as follows:
\begin{equation} \label{eq1}
\begin{gathered}
v_a=\frac{\mathcal{D}_a}{v_0+\sum_{a=1}^{M}{\mathcal{D}_a}}\\
\mathcal{D}_a=DSC(\tilde{l}_a,Seg_{inital})
\end{gathered}
\end{equation}
where $\mathcal{D}_a$ denotes the DSC between the current warped label $\tilde{l}_a$ and the initial segmentation. After getting all the weights $v_A$, the fusion of the obtained segmentation and co-registrations $W_A=\{Seg_{initial}, \;\widetilde{L}_A\}$ is performed. The fused result, which represents the output of the entire model, is as follows:
\begin{align}
    output=\sum_{i=1}^{M} v_i \cdot w_i\;; \;v_i \in V_A,w_i \in W_A
\end{align}

\begin{table*}[htbp]
\renewcommand\arraystretch {1.8}
\setlength{\tabcolsep}{3.2mm}{
\centering
\caption{Comparison of segmentation results with state-of-the-art segmentation networks for thyroid nodules on different datasets, including five comparison methods and ASTN with each method as the backbone, and the results of each metric are given in the form of mean ± standard deviation. }
\label{compare}
\begin{tabular}{c|cccc|cccc}
\hline
\multirow{2}{*}{\textbf{Method}} & \multicolumn{4}{c|}{\textbf{P1}}                                                    & \multicolumn{4}{c}{\textbf{M3}}                                                     \\ \cline{2-9} 
                                 & \textbf{DSC↑}      & \textbf{IOU↑}      & \textbf{HD↓}       & \textbf{ASSD↓}       & \textbf{DSC↑}      & \textbf{IOU↑}      & \textbf{HD↓}       & \textbf{ASSD↓}       \\ \hline
FDNSCL(2023)                     & 91.09±4.1          & 83.63±2.1          & 4.958±2.2          & 0.3823±0.22          & 76.17±6.5          & 61.51±3.4          & 14.78±2.6          & 0.8194±0.62          \\
\textbf{ASTN-FDNSCL}             & \textbf{91.88±2.6} & \textbf{84.97±1.3} & \textbf{4.846±1.5} & \textbf{0.1873±0.26} & \textbf{79.52±5.9} & \textbf{66.00±1.4} & \textbf{11.31±2.0} & \textbf{0.6564±0.41} \\ \hline
Transdeeplab(2022)               & 89.89±2.5          & 81.63±1.2          & 6.487±3.1          & \textbf{0.3238±0.28} & 75.19±6.6          & 60.24±3.4          & 11.68±8.5          & 0.8319±0.74          \\
\textbf{ASTN-Transdeeplab}       & \textbf{90.76±7.3} & \textbf{83.08±3.8} & \textbf{6.357±2.1} & 0.4630±0.44          & \textbf{83.15±3.3} & \textbf{71.16±1.7} & \textbf{10.58±2.1} & \textbf{0.7236±0.68} \\ \hline
SegNet(2017)                     & \textbf{93.22±3.5} & \textbf{87.30±1.8} & \textbf{4.849±2.5} & 0.1953±0.18          & 74.74±6.3          & 59.67±3.3          & 12.56±1.5          & 1.059±0.83           \\
\textbf{ASTN-SegNet}             & 92.15±3.2          & 85.44±1.6          & 5.194±3.7          & \textbf{0.1713±0.12} & \textbf{80.39±6.6} & \textbf{67.21±3.4} & \textbf{11.29±9.0} & \textbf{0.8682±0.58} \\ \hline
DeconvNet(2015)                  & 88.57±4.7          & 79.48±2.4          & \textbf{9.860±3.2} & 0.3627±0.29          & 68.46±3.4          & 52.05±1.7          & 25.62±8.6          & 1.679±0.84           \\
\textbf{ASTN-DeconvNet}          & \textbf{91.73±3.2} & \textbf{84.72±1.6} & 10.83±3.9          & \textbf{0.2176±0.28} & \textbf{72.21±4.2} & \textbf{56.51±2.1} & \textbf{16.06±6.2} & \textbf{1.068±0.57}  \\ \hline
Unet(2015)                       & 92.27±4.9          & 85.65±2.5          & 6.893±2.0          & 0.1493±0.15          & 72.78±5.1          & 57.21±2.6          & 36.41±9.3          & 2.457±1.7            \\
\textbf{ASTN-Unet}               & \textbf{92.57±3.6} & \textbf{86.18±1.8} & \textbf{6.026±2.8} & \textbf{0.1388±0.17} & \textbf{76.86±4.8} & \textbf{62.42±2.5} & \textbf{24.53±7.5} & \textbf{1.843±0.39}  \\ \hline
\end{tabular} }
\end{table*}

\section{Experiments and Results}
\subsection{Datasets}

We conducted experiments using the TUI dataset collected from collaborating hospital, comprising 11,360 images from the P1 device and 800 images from the M3 device. The P1 dataset was utilized to assess the co-registration and segmentation performance of the model, encompassing 4,796 benign nodule images and 6,564 malignant nodule images, all expertly annotated. To maintain class balance, we randomly selected 4,500 images from both benign and malignant nodule images, as the P1 training set, while the remaining 2,360 images served as the P1 test set. Furthermore, all images were resized to $(224,224)$ using bilinear interpolation. The M3 dataset consisted of 400 benign nodule images and 400 malignant nodule images, serving as the evaluation set for the generalization of ASTN. Moreover, the M3 dataset underwent the same preprocessing steps as the P1 dataset.

\subsection{Model Settings and Metrics}
The PyTorch framework was used for all model development. All codes were executed on an NVIDIA RTX3090 (with 24GB memory). We trained for 120 epochs, using RMSprop as the optimizer to optimize the model parameters. The learning rate of the Semantic Extraction network in the model was 0.00001, while the Deformation Fusion network had a learning rate of 0.0001. The learning rate decayed by 0.1 every 40 epochs. The batch size remained constant at 6. The dimension size of the latent space features of each ultrasound image was controlled at $1*1024$, and the dimension of DF was twice the size of the input image, i.e., $2*224*224$.

We validated the model's effectiveness through comparative and ablation experiments. These experiments verified the model's segmentation and registration effects from different angles. When comparing segmentation effects, we used common segmentation evaluation indicators. The Dice Similarity Coefficient (DSC, higher is better) measures the similarity between two images. It does this by calculating the ratio of the intersection size to the total pixels of predicted and real segmentation images. The Intersection over Union (IoU, higher is better), also known as the Jaccard coefficient, is similar to DSC, but IoU is less affected by changes in segmentation category size. The Hausdorff Distance (HD, lower is better) calculates the maximum distance between predicted segmentation and real segmentation images, capturing mismatched areas of segmentation and providing a more detailed accuracy assessment. The Average Symmetric Surface Distance (ASSD, lower is better) calculates the average distance between predicted segmentation and real segmentation images, similar to HD, but provides quantification of symmetric errors, which may be more informative for certain application scenarios.
There are described as:
\begin{equation}
\begin{gathered}
DSC = \frac{(2 * TP)} { (2 * TP + FP + FN)} \\
IoU =  \frac{TP }{ (TP + FP + FN)} \\
HD(P, Q) = max(hd(P, Q), hd(Q, P)) \\
ASSD = \frac12(ASD(P, Q) + ASD(Q, P))
\end{gathered}
\end{equation}
where $TP$ represents the number of true positive pixels (predicted as positive and actually positive), $FP$ represents the number of false positive pixels (predicted as positive but actually negative), and $FN$ represents the number of false negative pixels (predicted as negative but actually positive). $P$ and $Q$ respectively represent the predicted segmentation image and the actual segmentation image. $hd(P,Q)$ represents the shortest distance from each pixel in $P$ to $Q$, that is, the nearest neighbor distance. $ASD(P,Q)$ represents the average distance from each pixel in $P$ to $Q$, that is, the average distance between all pixels.

\subsection{Comparison Experiment}

\paragraph{Comparison with Segmentation Models}

To evaluate the precision and generality of ASTN, we employed common segmentation networks FC-DenseNet+SA+C-LSTM (FDNSCL)\cite{rondinella2023boosting}, Transdeeplab\cite{azad2022transdeeplab}, SegNet\cite{badrinarayanan2017segnet}, DeconvNet\cite{noh2015learning}, Unet\cite{ronneberger2015u} as the backbone, respectively, and compared the segmentation results with these networks on the P1 and M3 datasets, as summarized in TABLE~\ref{compare}. For a comprehensive evaluation, we trained each model on the P1 training set and then tested it on both the P1 test set and the M3 dataset. The evaluations are conducted from various perspectives using the commonly employed metrics mentioned above. When tested on the P1 dataset, which shares the same domain as the training data, the improvements in segmentation performance were marginal across all compared methods. ASTN showed a slight advantage in the DSC metric, especially when integrated with DeconvNet, Unet, FDNSCL, or Transdeeplab as the backbone. However, on the unseen domain M3, as indicated on the right side of TABLE~\ref{compare}, our method significantly outperforms existing methods in all metrics, with DSC and IoU exceeding the comparative methods by 5\% and 6.524\%, respectively. This demonstrates the superiority of our approach in thyroid nodule segmentation compared to all the alternative methods, particularly in the case of unseen domain, as shown in Fig.~\ref{bigfig}. Moreover, the robust performance of ASTN across various encoder-decoder architectures underscores its versatility and adaptability in different segmentation model frameworks.

\begin{table}[htbp]
\centering
\setlength{\tabcolsep}{5mm}{
\caption{Comparison of DSC results of different methods in co-registration stage and fusion stage}
\label{stage}
\begin{tabular}{c|c|cc}
\hline
\multirow{2}{*}{\textbf{Stage}} & \multirow{2}{*}{\textbf{Method}}     & \multicolumn{2}{c}{\textbf{DSC↑(\%)}}       \\ \cline{3-4} 
                        &                             & \textbf{P1}             & \textbf{M3}            \\ \hline
\multirow{5}{*}{\textbf{\uppercase\expandafter{\romannumeral1}$^{\mathrm{a}}$}
}      & SyNCC                       & 29.01
           & 36.57
          \\
                        & Voxelmorph                  & 54.92           & 52.06          \\
                        & hu                          & 40.18           & 40.68          \\
                        & U-ReSNet                    & 63.73           & 56.15          \\
                        & \textbf{ASTN w.o$^{\mathrm{c}}$ WLF(ours)} & \textbf{88.59}  & \textbf{71.38} \\ \hline
\multirow{3}{*}{\textbf{\uppercase\expandafter{\romannumeral2}$^{\mathrm{b}}$
}}      & MV                          & 89.8           & 73.95          \\
                        & STAPLE                      & 90.48          & 73.13          \\
                        & \textbf{ASTN(ours)}         & \textbf{92.57} & \textbf{76.86} \\ \hline
                        \multicolumn{4}{l}{$^{\mathrm{a}}$co-registration stage. $^{\mathrm{b}}$fusion stage. $^{\mathrm{c}}$w.o, without.}
\end{tabular} }
\end{table}

\begin{figure*}[t]
\centerline{\includegraphics[scale=0.25]{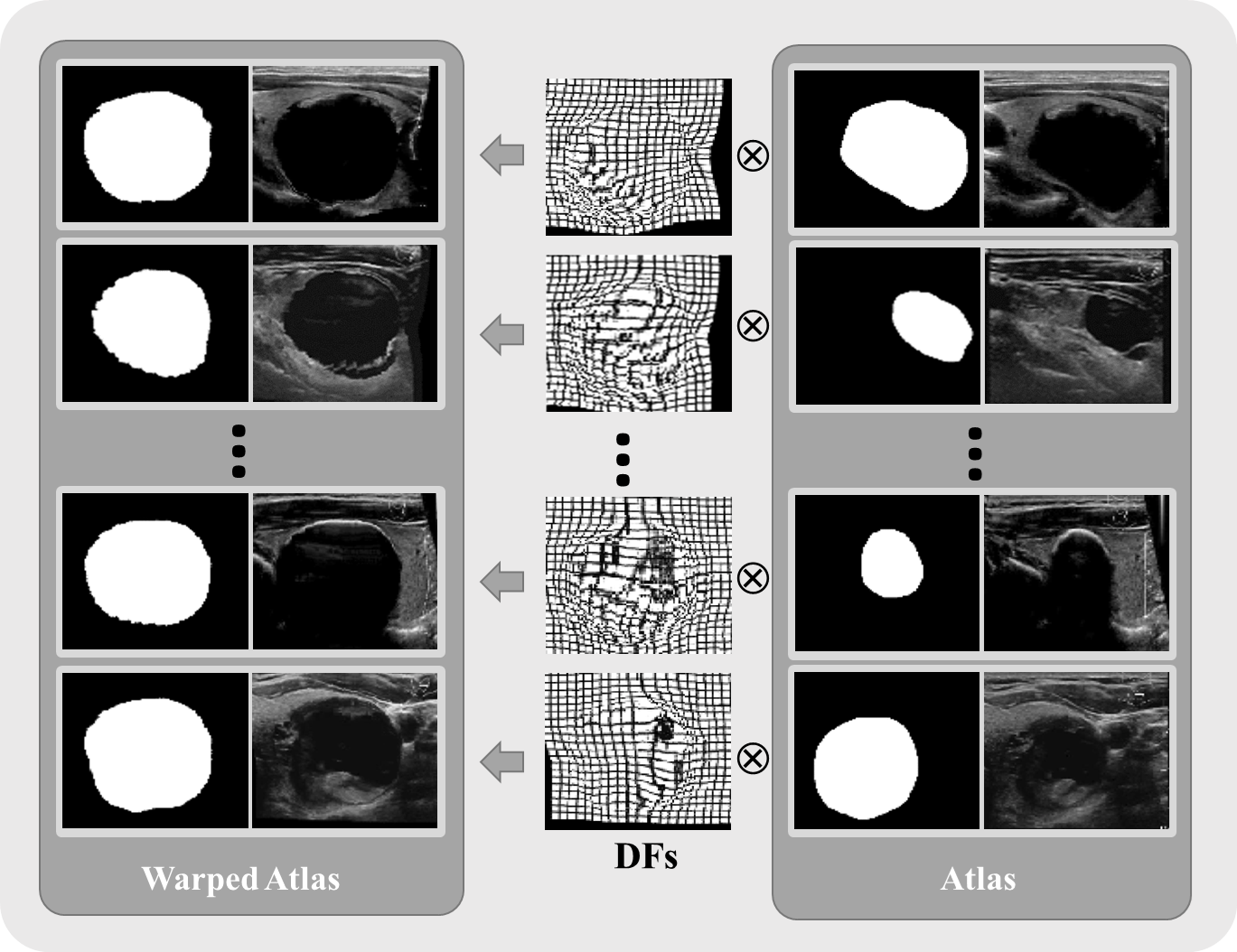}}
\caption{\textbf{Co-Registration.} To study the effectiveness of co-registration, the process of co-registering atlas to one of the target images is visualized. The atlas is shown on the right, the warped atlas on the left, and the corresponding deformation field (DF) for each element of the atlas is shown in the middle. The nodule area of the warped atlas is similar to the target, thereby substantiating the efficacy and rationality of our model.}
\label{register}
\end{figure*}

\paragraph{Comparison with co-Registration Models}In this segment of our experimental evaluation, we focused on assessing ASTN's co-registration performance in segmenting thyroid nodules. To isolate the impact of co-registration, we temporarily deactivated the Weighted Labels Fusion (WLF) component within ASTN. This allowed for a direct comparison between the warped labels generated by ASTN’s Spatial Transform component and the annotated segmentation labels provided by medical professionals. Additionally, we benchmarked ASTN against well-established co-registration methods, including VoxelMorphVoxelMorph\cite{balakrishnan2019voxelmorph}, SyNCC\cite{avants2009advanced}, Hu\cite{hu2018weakly}, and U-ResNet\cite{estienne2019u}. From the TABLE~\ref{stage} Stage \uppercase\expandafter{\romannumeral1}, it can be observed that prior to fusion, the co-registration DSC of ASTN already reaches 88.6\%, significantly outperforming the existing methods by an average DSC margin of over 15\%. To visually demonstrate the co-registration efficacy, we included in Fig.~\ref{register} a detailed representation of a specific segmentation process. This figure illustrates the atlas images and labels both before and after co-registration, showcasing the precise alignment of nodule areas from the atlas labels to the target. Notably, the ultrasound images reveal a marked closeness of nodule areas between the atlas and target post-co-registration, affirming the validity of our warped atlas labels and enhancing the interpretability of our method.

At the same time, this experiment shed light on the effectiveness of the WLF component in ASTN. The comparison of the network’s performance, with and without WLF, showed average DSC values of 92.57\% and 88.6\%, respectively, as recorded in TABLE~\ref{stage}. It demonstrates that WLF contributes to a roughly 4\% improvement in the final segmentation.

\paragraph{Comparison with Label Fusion Method} In this section, we compare two common label voting algorithms. The majority voting (MV)\cite{heckemann2006automatic} assumes equal fusion weights for all atlases, while the STAPLE\cite{warfield2004simultaneous} utilizes warped atlas labels to estimate the true probabilities of segmentation. For this comparison, we employ the Spatial Transform component output as the input for the voting algorithms. The results are presented in TABLE~\ref{stage} Stage \uppercase\expandafter{\romannumeral2}. On datasets P1 and M3, ASTN achieves higher DSC and HD results compared to MV and STAPLE methods. These findings demonstrate the efficacy of our warped label weight calculation method.

\subsection{Hyperparameter and Ablation study}
\begin{figure}[htbp]
\centerline{\includegraphics[scale=0.3]{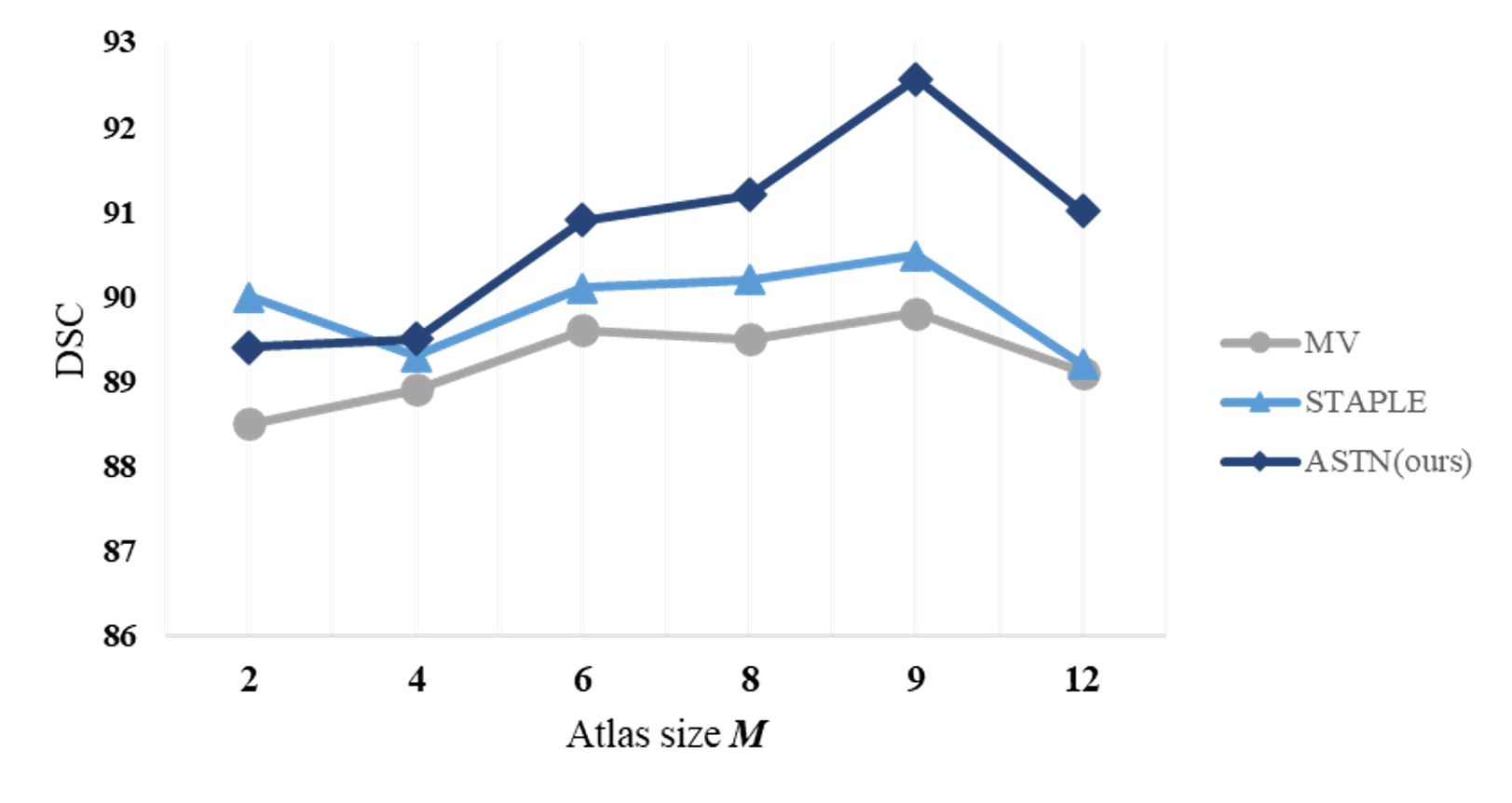}}
\caption{\textbf{The impact of varying atlas size $M$.} The horizontal axis represents the size of atlas, while the vertical axis represents the final segmentation DSC. Different colored curves correspond to distinct fusion Methods.}
\label{zhexian}
\end{figure}

\paragraph{Impact of Size of Atlas} We explored the impact of varying the size of the atlas within the RCS atlas selection algorithm on the P1. Experiments are performed on the label fusion algorithms of MV, STAPLE, and ours, with $M$ values of $\{2,4,6,8,9,12\}$, as shown in Fig.~\ref{zhexian}. Compared to the other two methods, the label fusion approach in ASTN is more sensitive to the number of atlas elements and achieves the optimal result when $M$ is 9.

\paragraph{Impact of RCS} While validating the efficacy of RCS, we conduct a comparative analysis on the influence of different atlas on co-registration and segmentation performance. Specifically, we keep the value of $M$ constant at 9 and compute the final segmentation DSC using two distinct atlases: one formed using the RCS and the other selected randomly. The results, as shown in TABLE~\ref{Ablation}, reveal that the atlas chosen based on RCS exhibited an average DSC of 92.57\%, whereas that for the randomly selected atlas is a mere 71.6\%. A subset of the atlas selected using the RCS on P1 is visually represented in Fig.~\ref{register}. Both the visual observations and quantitative results provide evidence that the atlas selected based on our method exhibits a uniform dispersion of nodules, indicating a high level of tolerance in nodule positioning during the co-registration process to target images.
\begin{table}[htbp]
\centering
\setlength{\tabcolsep}{8mm}{
\caption{Ablation study on ASTN}
\label{Ablation}
\begin{tabular}{c|cc}
\hline
\multirow{2}{*}{\textbf{Method}} & \multicolumn{2}{c}{\textbf{DSC↑(\%)}} \\ \cline{2-3} 
                                 & \textbf{P1}     & \textbf{M3}     \\ \hline
ASTN w.o$^{\mathrm{a}}$ RCS                     & 71.63           & 42.83           \\
ASTN w.o DecOfSeg$^{\mathrm{b}}$                & 61.4            & 51.09           \\
\textbf{ASTN}                    & \textbf{92.57}  & \textbf{76.86}  \\ \hline
\multicolumn{3}{l}{$^{\mathrm{a}}$w.o, without. $^{\mathrm{b}}$DecOfSeg, decoder of segmentation}
\end{tabular}}
\end{table}

\paragraph{Impact of Decoder of Seg} To validate the contribution of the segmentation network decoder, we train the ASTN without it. This modification led to performance metrics on par with mainstream co-registration methods, yielding an average Dice Similarity Coefficient (DSC) of 61.4\%. Since the original segmentation decoder is removed, joint training of segmentation and co-registration is not possible, and the absence of the segmentation loss function $L_{sim}$ in network optimization meant that the encoder is not constrained to focus on the semantic information of the nodule area. Consequently, the accuracy of the resulting model without the decoder was on par with standard co-registration methods, but markedly inferior to our full ASTN framework.

\section{Conclusion}
In this paper, we propose an end-to-end deep neural network, ASTN, and a new atlas selection algorithm to address the problem of thyroid ultrasound image nodule segmentation. ASTN is distinguished by its ability to confine the registration network’s operation to the lesion area by extracting latent space features. It also completes segmentation by better preserving the biological anatomical structure with reference to the prior information in the atlas. Furthermore, our innovative atlas selection algorithm, the Regional Correlation Score, optimizes the distribution of elements within the atlas, thereby enhancing registration accuracy. The new label fusion strategy combines information from the target at multiple scales, improving the tolerance for registration and enhancing segmentation quality. Our comparative studies with recent methodologies and classical segmentation networks reveal significant improvements in DSC, IoU, HD, and ASSD metrics by 4.958\%, 6.524\%, 5.456\%, and 3.374\% respectively, across different vendor ultrasound image datasets. These advancements are primarily a result of separately optimizing feature extraction and spatial transformation processes using distinct loss functions, achieving localization of the nodule area before deformation, and filtering out redundant information from other body parts in the image. Once registration focuses more on the lesion area, the overall accuracy of registration and segmentation results will significantly improve.

Despite these promising results, the clinical applicability of ASTN still needs to be validated through more different datasets, especially datasets of different imaging modalities, such as X-ray, CT, etc. This future direction calls for extended collaborations with medical institutions or access to diverse public datasets. Additionally, while our atlas selection algorithm performs well with single-nodule thyroid ultrasound datasets, the presence of multiple nodules introduces new complexities. In the future, we will focus on proposing new atlas construction methods to solve this problem, such as replacing the fixed atlas with an atlas that can be updated with the model, which could be another potential direction for ultrasound image segmentation.

\section*{Declaration of Competing Interest}
The authors declare that they have no known competing financial interests or personal relationships that could have appeared to influence the work reported in this paper.
\section*{Data availability}
The authors do not have permission to share data.
\section*{Acknowledgments}
The authors of this paper would like to thank Dr. Xi Wei for providing data support from the Tianjin Medical University Cancer Institute and Hospital for the project research.

\bibliographystyle{ieeetr}
\bibliography{main}

\end{document}